\documentstyle[preprint,aps]{revtex}
\newcommand \sqr {\sqrt{3}\times\sqrt{3}}
\begin{document}

\title{ First Principles Calculations of Charge and Spin Density 
	Waves of $\sqrt{3}$-Adsorbates on Semiconductors }

\author{ Sandro Scandolo$^{1,2,3}$, Francesco Ancilotto$^{4,3,2}$, 
         Guido L. Chiarotti$^{2,3}$, Giuseppe Santoro$^{2,3}$,
         Stefano Serra$^{2,3}$, and Erio Tosatti$^{1,2,3}$,\\
       }

\address{
	 $1)$ International Centre for Theoretical Physics (ICTP),
	      I-34014 Trieste (Italy) \\
         $2)$ International School for Advanced Studies (SISSA), 
              Via Beirut 4, I-34014  Trieste (Italy)\\
	 $3)$ Istituto Nazionale di Fisica della Materia (INFM) \\
         $4)$ Dip. di Fisica, Via Marzolo 8, Universit\'a di Padova (Italy)
        }
\maketitle

\begin{abstract}
We present ab-initio electronic structure results on the surface
of $\sqr$ adsorbates. In particular, we address the issue of
metal-insulator instabilities, charge-density-waves
(CDWs) or spin-density-waves (SDWs), driven by partly filled surface 
states and their 2D Fermi surface, and/or by the onset of magnetic 
instabilities. 
The focus is both on the newly discovered commensurate CDW transitions
in the Pb/Ge(111) and Sn/Ge(111) structures, and on the puzzling 
semiconducting behavior of the Pb/Ge(111), K/Si(111):B and SiC(0001) surfaces.
In all cases, the main factor driving the instability appears to be
an extremely narrow surface state band. 
We have carried out so far preliminary calculations for the Si/Si(111) 
surface, chosen as our model system, within the gradient corrected local 
density (LDA+GC) and local spin density (LSD+GC) approximations, with the 
aim of understanding the possible interplay between 2D Fermi surface and 
electron correlations in the surface + adsorbate system. 
Our spin-unrestricted results show that the $\sqr$ paramagnetic surface is 
unstable towards a commensurate density wave with periodicity $3\times 3$ 
and magnetization 1/3.   
 
\end{abstract}

\section{Introduction}

Semiconductor surfaces are generally believed to belong to the
realm of solid state problems where electron correlations are not
too important. Very recently, however, a new phase transition has been 
observed on Ge(111), when this surface is covered with 1/3 of a 
monolayer of Pb adatoms~\cite{carpinelli}.
The $\sqr$ room-temperature $\alpha$-phase of Pb/Ge(111) 
has been observed to transform reversibly into a new phase, with $3\times 3$ 
periodicity, below 250 K, with noticeable changes in the electronic structure 
~\cite{modesti,avila}. 
The same phenomenon has been reported for
Ge(111) covered with Sn~\cite{plummer,modesti2}, at a slightly lower (210 K) 
transition temperature. Reconstructions are ubiquitous in semiconductor 
surfaces, but the finding of a continuous, reversible phase transition 
as a function of temperature, plus the close agreement of the
low-temperature surface periodicity with a calculated Fermi surface``nesting'' 
vector of Pb/Ge(111) is new, and suggested that the transition 
might be a clear example of surface-state driven charge-density wave 
\cite{CDW}. However, there are clear problems with this picture in its
simplest form. First, we note, the true nesting cannot be very good,
given one electron/adatom. Second, the $3\times 3$ state still contains 
an odd electron number/cell, and should be very metallic, while,
strikingly, EELS evidence suggests, at least for Pb/Ge(111), a small but 
finite gap or pseudogap ~\cite{carpinelli}. 
Much larger and clearer insulating gaps have 
moreover been recently found on other isoelectronic  $\sqr$ surfaces, 
such as K/Si(111):B ~\cite{weitering}, and Si-terminated SiC(0001)~\cite{sic}, 
where electron counting arguments would similarly predict band metallicity. 
No structural data are available, however, for K/Si(111):B and SiC(0001).

In all the above systems, the surface band arising from the half-filled 
adatom dangling bond orbital (pointing outward from the (111) surface)
displays  a weak dispersion over the surface Brillouin Zone (SBZ), with
typical calculated bandwidths ranging between 0.35 in SiC(0001) 
and 0.5 $\div$ 0.6 eV in Pb- or Sn- covered Ge(111). The adatoms are in
fact sitting very widely apart, the only source of electron hopping between 
them requiring higher order hops through the back bonds and the substrate. 
If such a small bandwith is compared with
a relatively large on-site Coulomb repulsion for two electrons when
occupying the same dangling bond
orbital, or even two neighboring ones, then it becomes suggestive
to suppose that strong correlations might play a role in determining the 
true electronic surface ground state, along with the detailed equilibrium 
atomic geometry~\cite{pippo}. 
In particular, K/Si(111):B and SiC(0001) would correspond
in this picture to Mott-Hubbard insulators.

Strongly correlated electron systems are, strictly speaking, not
tractable with effectively one electron methods, such as Hartree-
Fock, or Local Density (LDA) approximations. This is all the more
lamentable, since these approximations are very good at describing
the basic bulk chemistry of these semiconductors, and we have 
nothing of comparable simplicity and accuracy to replace them, once
their validity is impaired by correlations, as it is in Mott insulators.

There is however at least one well-known trick one can resort to,
which may be successful, provided one does not forget its deep
limitations. The idea is to make use of the fact that Mott insulators 
are dominated by magnetic correlations. If we extend the one-electron
methods to include the possibility of developing static magnetic
order parameters, we can hope to recapture, if not the full
strongly correlated state, at least a mimic of its local aspects,
which may be energetically close enough to the truth ~\cite{fulde}.
Hence, a further important step to move before abandoning these 
systems, is to switch from restricted Hartree-Fock to unrestricted
Hartree-Fock, or from Local Density to Local Spin Density (LSDA)
approximations. Here, long-range magnetic order is permitted, and 
can both lower the
energy and yield insulating states in a half-filled band. If that
should happen, it will of course not necessarily follow that the true system
must have long-range magnetic order, since in the Mott phenomenon
the insulator precedes the magnet, and not viceversa. All the same,
such a calculation is nonetheless quantitative (i.e., variational), and it 
may in fact teach us a great deal.

In this work we present a preliminary study of the basic physics 
of these systems, conducted by comparing LDA with LSDA calculations
of a model  $\sqr$ surface.  
We focus our interest to Si/Si(111) $\sqr$, chosen as a prototype case, 
for two reasons: {\it (i)} among the structures displaying ``unconventional''
behavior, the K/Si(111):B surface (which should closely resemble Si/Si(111) )
is the one where correlations are larger, surpassed only by SiC(0001). 
{\it (ii)} the availability of accurate theoretical and experimental 
structural data
on the atomic configuration of Si/Si(111) allows us to focus on the electronic 
issues rather than on the more complex interplay between electronic and atomic
degrees of freedom, that will instead form the subject of future 
investigations.

We performed extensive electronic structure calculations for this surface,
both in the local density approximation
(LDA) and in the local spin density approximation (LSDA), and we supplemented 
both types of calculations with gradient corrections (GC) to the energy 
functional. We employed a plane-wave basis set with 9 Ry energy cutoff, 
and we used a maximum of about 1000 k-points to sample the full SBZ of the 
$\sqr$ phase. In Section II we discuss the properties of the undistorted 
Si/Si(111) $\sqr$ surface, and show that in fact correlations 
do make this surface 
unstable towards a magnetic state. We also show that a state where the surface 
band is 2/3 filled with spin-up electrons and 1/3 with spin-down electrons 
(i.e. where the magnetization $M=1/3$), displays a much stronger tendency to
develop a $3\times 3$ density wave than the $M=0$ paramagnetic case. In 
this case, it turns out, nesting is made strong -- nearly perfect -- 
by the fractional magnetization. 
In Section III we present preliminary results of a calculation performed 
with a $3\times 3$ surface unit cell, where we confirm that a state with
$M=1/3$ and $3\times 3$  charge and spin periodicity develops. 
Conclusions and prospects for future work will be given in Section IV.

\section{The unreconstructed surface}

The Si/Si(111) $\sqr$ unreconstructed surface was modeled with a slab 
containing two Si bilayers plus a Si adatom that was placed in the $T_4$ 
position of the upper surface. All atomic positions were fixed to the values
calculated by Northrup ~\cite{northrup}. The bottom surface (fourth atomic 
layer), which is not planar, was saturated with H atoms. 
With this choice, our calculated 
Hellmann-Feynman forces were smaller than 0.05 eV/\AA\ on all the atoms 
of the slab, indicating a reasonably stable starting state. We 
initially carried out a paramagnetic calculation (LDA$+$GC)
in order to determine the surface band dispersion and the 
Fermi surface that originates from half-filling this surface band.

The surface band was located in the middle of the bulk-projected 
energy gap, and had a width of about 0.6 eV.
As stated in the introduction, the nesting vector connecting two 
parallel portions of the Fermi 
surface (see Fig. 1) is clearly larger than the value required to justify
a commensurate charge-density wave with $3\times 3$ periodicity, which
corresponds to twice the $\Gamma \to M_{3}$ distance in Fig. 1.
This is in agreement with recent calculations reported for the Sn/Ge(111) 
~\cite{plummer}
surface, and disagrees with earlier claims for Pb/Ge(111)
~\cite{carpinelli}.
Since a low temperature transition to a $3\times 3$ phase has been 
observed on both Sn/Ge(111) and Pb/Ge(111), the role of a Fermi surface 
nesting as the driving force of the transition remains to be clarified.
A possible low temperature transition to a $3\times 3$ phase has not been 
investigated experimentally for the K/Si(111):B system, or for SiC(0001). 
However, our paramagnetic calculation suggests discarding the nesting argument 
in all of these systems.

The above calculations have been carried out within the LDA$+$GC 
approximation, and are thus inclusive of electron correlations only at a 
mean field level. This approximation is well-known to be sufficiently accurate
only for those systems where the band width $W$ is not too small with
respect to the on-site Coulomb repulsion $U$, and nearest-neighbor 
repulsion $V$. A crude estimate of $U$ and of $V$ in 
our surface can be obtained by constructing the Wannier function associated
with the surface band (see Fig. 2), 
evaluating its Coulomb integrals $U_\circ$, and $V_\circ$, and 
screening them by the electronic response of semi-infinite bulk Si,
so that $U\simeq U_\circ \frac{2}{\varepsilon +1} \simeq U_\circ/6$,
and the same for $V$.
Evaluation of $U_\circ$ for the Wannier function of Fig. 2 
gives $U_\circ = 3.6$ eV, $V_\circ = 1.8$ eV and 
$U \simeq 0.6$ eV, $V \simeq 0.3$ to be compared with a band width 
$W\simeq 0.5$ eV. 

This shows that correlations will indeed be strong, as was surmised.
This estimate of $U$, for example, suggests that the system violates 
the Stoner criterion for the stability of the paramagnetic state. 
In fact, estimating a density of states at the Fermi level as $n(E_F) 
\simeq 2/W$ (this estimate holds for a flat density of states of
width $W$, the {\em calculated} $n(E_F)$ is slightly larger), we get 
$n(E_F) U \simeq 2.4 > 1$. In other words,
the Stoner criterion strongly suggests that the paramagnetic state considered
so far is {\em unstable} with respect to a magnetic state whose character
remains to be determined. Because of this, it will be instructive
to switch, as explained in the Introduction, to LSDA calculations.

We have carried out a spin-polarized calculation of Si/Si(111), 
within the LSD$+$GC approximation. For simplicity, we only considered the state
with magnetization $M=1/3$, obtained by filling the surface band with
2/3 of spin-up electrons and 1/3 of spin-down electrons. 
In agreement with what suggested by the Stoner criterion,
we found that the $M=1/3$ state is favored, by about 10 meV/adatom,
with respect to the unpolarized case. While this calculation is only 
representative of the fact that a magnetic instability has to set in,
the ``true'' ground state of the system being still to be determined, 
it is suggestive to observe that the Fermi surface for spin-up electrons
in the $M=1/3$ calculation (see Fig. 3) now displays an exceedingly 
strong nesting in 
correspondence to the  $3\times 3$ reciprocal vector. In other words, 
although the $M=1/3$ state may not be the ground state for a $\sqr$ 
periodicity, such a magnetization could be strongly stabilized by 
a concomitant $3\times 3$ density wave.

\section{The $3\times 3$ distorted surface}

	In order to verify the hypothesis that a $3\times 3$ density wave can 
stabilize a $M=1/3$ state, we carried out a preliminary study of 
the electronic ground state in a $3\times 3$ surface unit cell, 
within the LSD$+$GC. In this calculation we did not impose a value of 
$M=1/3$, but we allowed the magnetization to reach its optimum value in
a self-consistent manner. 
	The resulting magnetic moment spontaneously converged to 
$M=1/3$. At the same time, the system
developed a spin density wave such that one of the three adatoms was
mainly occupied by spin-down electrons, while the remaining two adatoms
were mainly of spin-up character. 
This {\em spin} density wave was accompanied by a very small 
{\em charge} density 
wave ($\Delta \rho / \rho \sim 10^{-2}$ in the surface region),
so that the total charge approximately 
preserved the ``unreconstructed'' $\sqr$ 
periodicity, at variance with what is suggested by the STM data on 
Sn/Ge(111) and Pb/Ge(111).
The resulting band structure (see Fig. 4) was semiconducting,
and developed an indirect gap of about 0.2 eV, with an average direct 
gap of about 0.5 eV . 

\section{Discussion and prospects}

Within LSDA, we have obtained a surface energy lowering in going
from a paramagnetic, metallic  $\sqr$ state, to a magnetic, SDW
(and in principle also CDW) state with $3\times 3$ periodicity and an 
insulating gap.
This is therefore a natural candidate for the ground state of  K/Si(111):B. 
Because of the limitations discussed above, and also because of
Mermin-Wagner's theorem, it remains unclear whether long-range magnetic
order could ever really develop on this surface. 
Among other things, a knowledge of spin-orbit coupling and of 
magnetic anisotropy will be needed. Nonetheless,
we believe we have obtained a static picture of what the local
correlations should be. More work is now in progress to extend this
work in order to get a more detailed picture, as well as understanding
to what extent correlations, even if surely weaker, might play a role
also on Pb/Ge(111) and Sn/Ge(111).  
Inclusion of lattice distorsions will also be considered, along with 
the possibility of spin noncollinearity, where the $M=1/3$ might 
eventually turn into a $120^\circ$ magnetic structure~\cite{pippo}.

\section{Acknowledgements}

We acknowledge financial support from INFM, through Project ``LOTUS'',
and discussions with S. Modesti.

\figure{
\caption{Fermi surface obtained by half filling the surface band of 
	 of Si/Si(111); the Brillouin zone of the $3\times 3$ 
 	 surface is also reported.}
}

\figure{
\caption{Density contours of the Wannier function associated with
	 the Si/Si(111) surface band: dots correspond
	 to atomic positions}
}

\figure{
\caption{Fermi surface obtained from the surface band of Si/Si(111)
	with a fractional filling of 2/3.
	 The Brillouin zone of the $3\times 3$ surface is also reported.
	 Notice the strong nesting properties of the Fermi surface.}
}

\figure{
\caption{ Surface band structure of the 3x3 reconstructed surface,
	  calculated in the LSD approximation. 
	  Solid line: spin-up electrons; dotted line: spin-down electrons.
	  Shaded regions: projected bulk bands.}
}

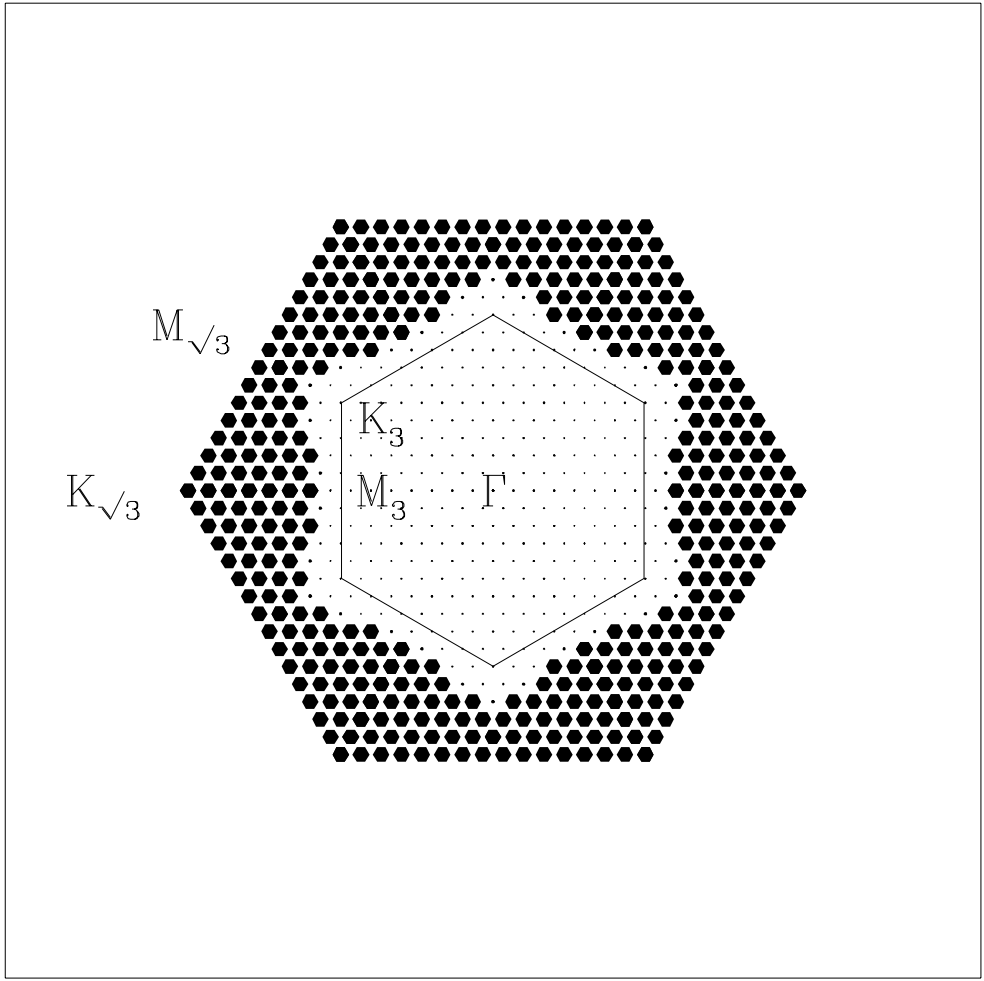
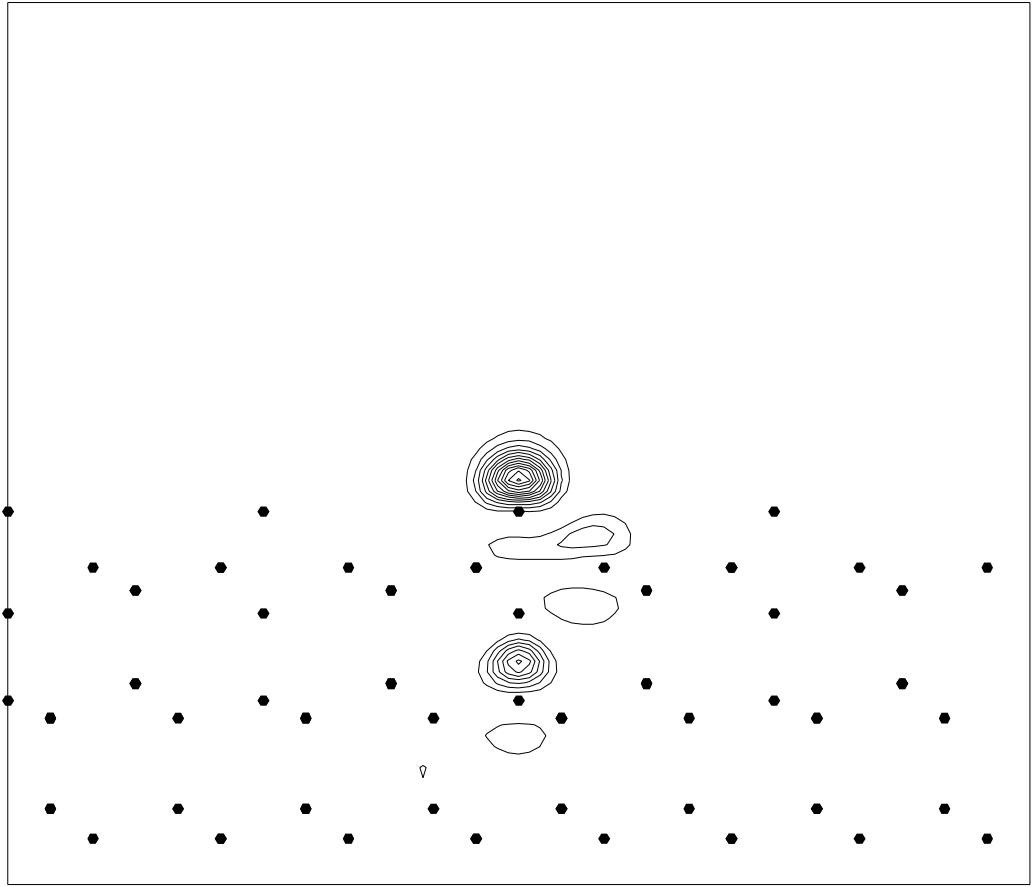
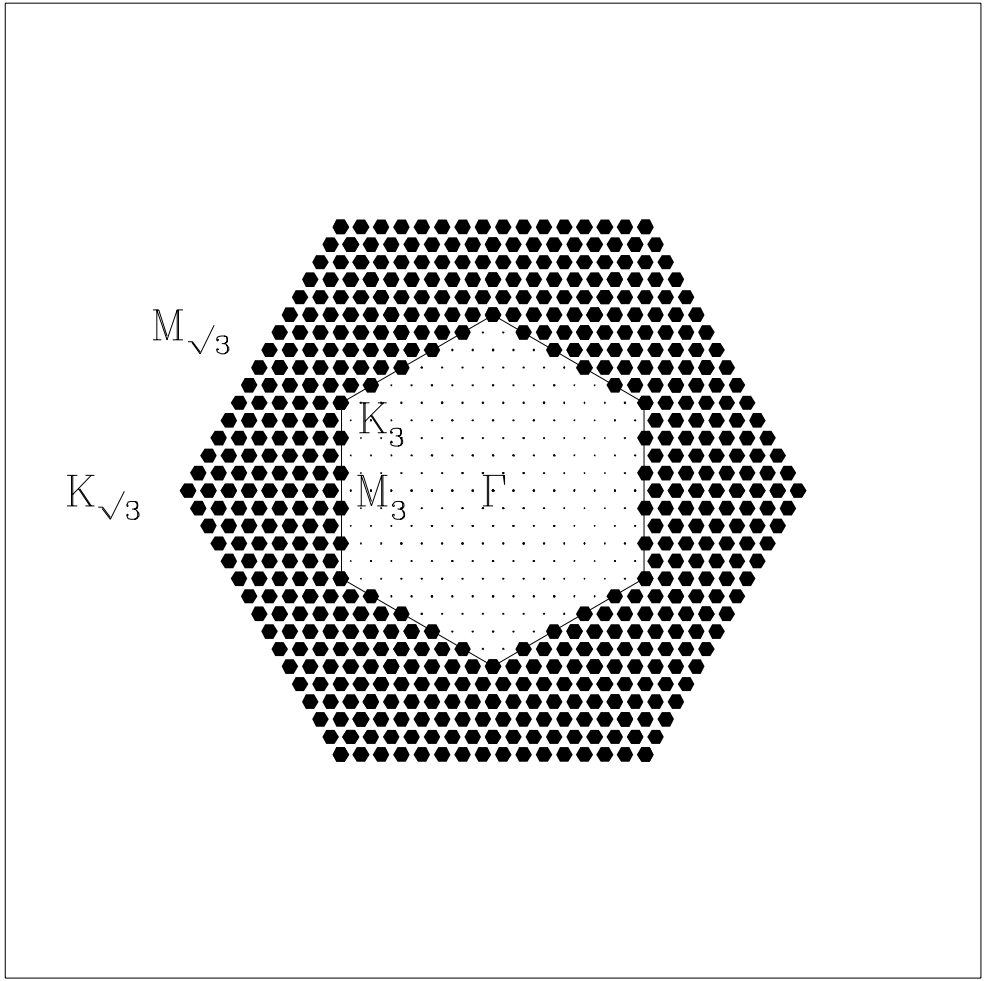
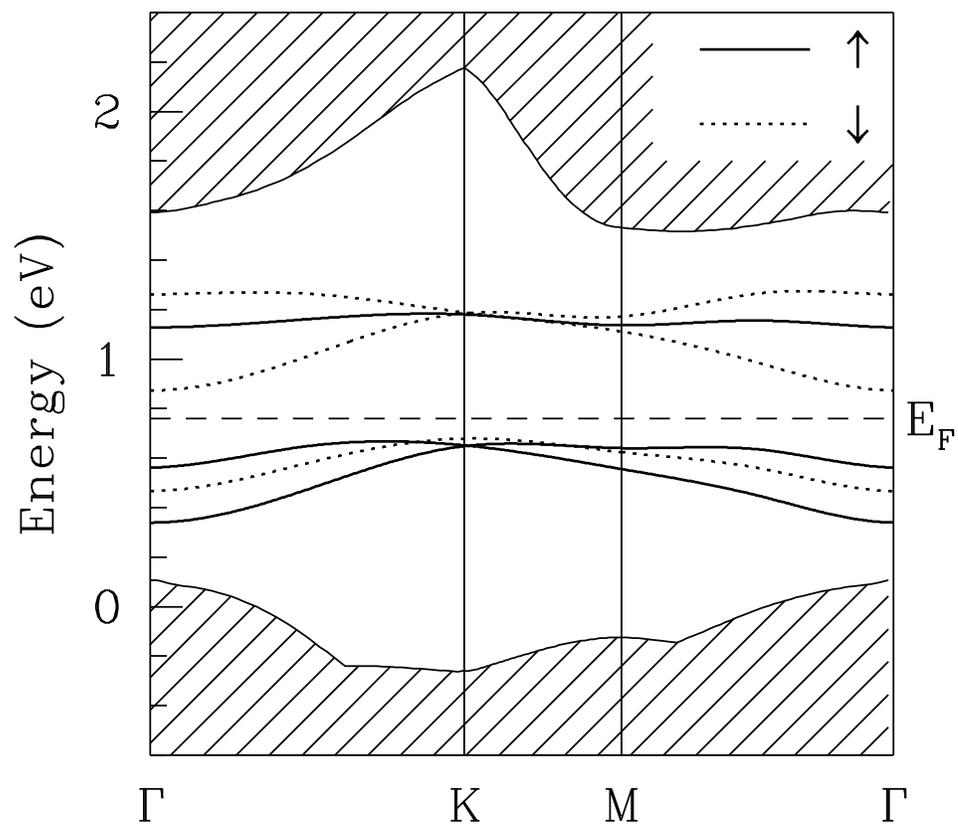
\end{document}